\documentclass{jltp}

\usepackage{graphicx,psfrag} % uncomment this line to include the graphicx package

\title{Random-Matrix Theory: \\
    Distribution of Mesoscopic Supercurrents through a Chaotic Cavity}
\author{Markus Garst$^{1}$ and Thilo Kopp$^{2}$}

\address{
$^1$ TKM, Universit\"at Karlsruhe, D--76128 Karlsruhe, Germany\\
$^{2}$ EP VI, EKM, Universit\"at Augsburg, D-86135 Augsburg, Germany\\}

\runninghead{M. Garst and T. Kopp}{RMT: Distribution of Mesoscopic Supercurrents \dots}
\begin{document}

\maketitle

\begin{abstract}
  We investigate the distribution of the supercurrent through a
  chaotic quantum dot which is strongly coupled to two superconductors
  when the Thouless energy is large compared to the superconducting
  energy gap. The distribution function of the critical currents
  $\rho(I_c)$ is known to be Gaussian in the limit of large channel
  number, $N\rightarrow\infty$. For $N=1$, we present an analytical
  low-temperature expression for this distribution function, valid
  both in the presence and in the absence of time-reversal symmetry.
  It connects directly the distribution of transmission coefficients
  to the distribution of critical currents. The case of arbitrary
  channel number ($N\ge2$) is discussed numerically, and for small
  critical currents analytically.

PACS numbers: 73.20.Dx, 74.50.+r, 74.60.Jg, 74.80.Fp, 73.20.Dx, 85.25.Cp.
\end{abstract}

\section{INTRODUCTION}
Electronic transport through mesoscopic structures with system size
$L$ smaller than the electron phase relaxation length $L_{\phi}$ has
become a unique ``playground'' to study quantum interference
effects.\cite{altshuler91,imry} An intriguing phenomenon which
convincingly presents the essence of quantum interference in diffusive
electronic transport are the universal conductance fluctuations (UCF).
Their size is of the order of $e^2/h$, independent of sample geometry
or the details of the realization of disorder, provided that the
electronic system stays in the diffusive transport regime. UCF were
first considered in a diagrammatic approach,\cite{altshuler85,lee} and
then soon discussed within random-matrix theory
(RMT)\cite{muttalib,mello} which Imry\cite{imry86} had proposed to
apply for this class of problems. UCF were observed as magneto-finger-prints
in mesoscopic wires (see the measurement of conductance fluctuations
in gold wires by Washburn and Webb\cite{washburn86}).

The superconducting analogue of UCF in normal conducting wires are
supercurrent fluctuations in chaotic SNS-Josephson point contacts,
as put forward by Beenakker.\cite{beenakker91,beenakker92} The
sample-to-sample fluctuations of the critical current $I_c$ depend only
on the gap in the single particle spectrum:
\begin{equation}\label{rms}
{\rm rms}\; I_c \equiv\, \sqrt{\langle I_c^2\rangle -\langle I_c\rangle^2}\,=
   \, c\, {e\Delta\over\hbar}
\end{equation}
with a numerical coefficient $c$ of order unity. Here, $\Delta$ is the
bulk gap, and the junction extension is assumed to be small compared
to the coherence length. Again, the fluctuations are universal in the
sense that they do not depend on junction parameters like, for example,
the degree of disorder or the junction geometry. These universal
supercurrent fluctuations still have to be verified experimentally.

The realization depends on several conditions. First of all, the
dynamics of the junction should be chaotic either due to impurities or
boundary scattering. For example, for a diffusive junction, the
junction length $L$ (which is to be taken as the separation of SN
interfaces) has to be much larger than the mean free path $l$ in the
normal-conducting part of the junction, $l \ll L$.  Secondly, to
ensure universality, the superconducting energy has to be the only low
energy scale. This amounts to the criterion that the Thouless energy
$E_c=\hbar/\tau_{d}$ is much larger than the maximum gap in the energy
spectrum, $E_c\gg\Delta$, where $\tau_d$ is the mean dwell time of
electrons or holes in the junction. For a diffusive junction, this
condition of a short dwell-time is fulfilled if $L$ is much smaller
than the coherence length, $L\ll\xi$, where $\xi=(\xi_o l)^{1/2}$ is
the diffusive coherence length, $\xi_o=\hbar
v_{\mbox{\tiny{F}}}/(\pi\Delta)$, and $v_{\mbox{\tiny{F}}}$ is the
Fermi velocity.  Moreover, the number $N$ of transverse modes at the
Fermi level which propagate through the junction, has to be large to
allow an expansion in the ``dimensionless inverse conductance''
$(2e^2/h)/\langle G \rangle$ and thus to prove that the numerical
coefficient $c$ in Eq.~(\ref{rms}) is indeed independent of
$N$.\cite{beenakker92}

Physically, (a) multiple Andreev reflection at the
interfaces,\cite{andreev} and (b) the statistics of the transmission
eigenvalues lead to the universal fluctuations, Eq.~(\ref{rms}).
Thereby, (a) is reponsible for the formation of current carrying bound
(Andreev-) states in the junction. In the short dwell-time regime,
only one Andreev state per channel exists. As a consequence, the
supercurrent $I(\phi)$ is a linear statistic\cite{footnote1} on the
normal-region transmission coefficients $T_n$ $(1\leq n\leq N)$,
independent of all junction parameters.  However, the critical current
$I_c\equiv \max I(\phi) \equiv I(\phi_c)$ is in general not a linear
statistic on $T_n$ since the phase $\phi_c$ at which the supercurrent
is maximal depends on all the transmission coefficients.  Beenakker
showed that $I(\phi_c)$ approximately obeys a linear statistic to
within a correction of order $1/N$, where $N$ is the number of
channels. From the statistical properties of the transmission
eigenvalues $T_n$ it is known that the fluctuations of a linear
statistic are of order unity and thus are universal, Eq. ~(\ref{rms}).

In this work we consider a Josephson junction where a chaotic quantum
dot is strongly coupled to two superconductors. It is known that the
scattering matrix of the normal conducting dot in this case belongs to
the ``circular unitary ensemble'' (CUE) of RMT.\cite{beenakker97} The
case of a weakly coupled junction was considered
elsewhere.\cite{brouwer97} We are especially interested in the
fluctuations of the supercurrent if the number of conducting channels
in the normal chaotic region, $N$, is small.  This implies that the
critical current does not obey a linear statistic.  Nevertheless, we
expect that the sample-to-sample fluctuations are still determined by
the relation Eq.~(\ref{rms}).
%since for $E_c \gg \Delta$ the only low energy scale stays $\Delta$.  
However, the
prefactor $c$ will now be weakly dependent on the number of channels
$N$.  More relevant is the question of how the full distribution of
critical currents, $\rho(I_c)$, will behave when we transit from a
situation with no level repulsion ($N=1$) to the scenarios with level
repulsion ($N\ge2$). Certainly, the two situations differ
qualitatively. For $N\rightarrow\infty$ the distribution is a
Gaussian.\cite{politzer} 
%From the central limit theorem one would
%expect fluctuations of the order of $\sqrt{N}$, however, the variance
%is of the order of unity due to a strong suppression of fluctuations
%by level repulsion.  
For a single-channel junction, the literature
does not provide a numerical or analytical result, to our knowledge.
The missing level repulsion will move the center of the distribution
to a lower value of $I_c$, the details of which should depend on the
chosen ensemble.  Finally, the case of a small number of channels is
intriguing because it will interpolate between these two qualitatively
incompatible situations --- no level repulsion versus repulsion
between many levels.

Although a Josephson junction with these properties has not yet
been realized, neither for a small nor for a large number of channels,
the microfabrication of point contacts with a small number of channels
is already feasible (for example, see Refs.~\onlinecite{scheer00,goffman,koops}); 
and findings like the conductance quantization, related to the 
Friedel sum rule,\cite{kirchner} open new
possibilities for the research of correlation effects in nano-scale
structures. Chaotic point contacts are now within reach of
experiments. The investigation of the interference of diffusion and
Coulomb interaction should soon be possible, and we expect new
exciting physics to emerge in these systems.

\section{MODEL}

In the recent 15 years, RMT has become a tool with which electronic
transport through disordered mesoscopic systems can be studied very
efficiently. It builds on the statistical independence of the elements
of, e.g.\, the scattering matrix and directly implements the symmetry of
this matrix.  As an excellent review by Beenakker\cite{beenakker97} on
this topic is available, we will only motivate some of the basic
results in this section. We are going to apply RMT to the normal
region of the SNS contact. The model which we use for the SNS contact
was suggested in Ref.~\onlinecite{beenakker92}. However, we will
extend the previous considerations to heterogeneous contacts.

\subsection{Random Matrix Theory for the Chaotic Region}

The critical current through an SNS-junction is determined by a
specific realization of a random potential in the normal region which
is the origin of the mesoscopic fluctuations. Interaction effects
between the electrons in the normal metal are not taken into account,
neither is spin-orbit coupling.

\begin{figure}
\vspace*{1cm}
\begin{center}
    \psfrag{Supraleiter 1}{\small $\!\!\!\!\!\!$superconductor 1}
    \psfrag{Supraleiter 2}{\small $\!\!\!\!\!\!\!\!\!\!\!$superconductor 2}
    \psfrag{Normalleiter}{\small $\!\!\!\!\!\!\!\!\!\!\!$normal conducting}
    \psfrag{Unordnungs-}{\small $\,\,\,$chaotic}
    \psfrag{bereich}{\small region}
    \psfrag{z1}{$z_1$}
    \psfrag{z2}{$z_2$}
    \psfrag{a1}{\small $\!\!\!\! a^+_e$}
    \psfrag{a2}{\small $\!\!\!\! a^-_e$}
    \psfrag{a3}{\small $\!\!\!\! a^-_h$}
    \psfrag{a4}{\small $\!\!\!\! a^+_h$} 
    \psfrag{b1}{\small $b^+_e$}
    \psfrag{b2}{\small $b^-_e$}
    \psfrag{b3}{\small $b^-_h$}
    \psfrag{b4}{\small $b^+_h$}
    \psfrag{c1}{\small $\!\!\! c^+_e$}
    \psfrag{c2}{\small $\!\!\! c^-_e$}
    \psfrag{c3}{\small $\!\!\! c^-_h$}
    \psfrag{c4}{\small $\!\!\! c^+_h$} 
    \psfrag{d1}{\small $d^+_e$}
    \psfrag{d2}{\small $d^-_e$}
    \psfrag{d3}{\small $d^-_h$}
    \psfrag{d4}{\small $d^+_h$}
     \leavevmode
\includegraphics*[width=4.8in,height=2.0in]{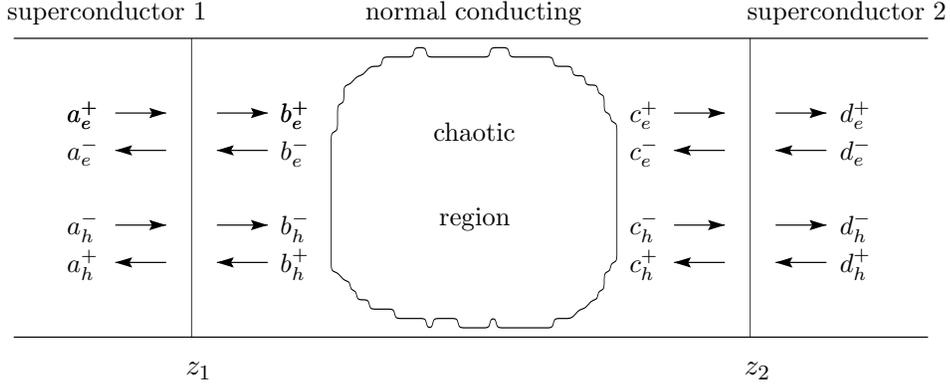} 
\end{center}
\caption{Junction geometry and definition of scattering amplitudes for 
particles ($e$) and holes ($h$).}
\label{fig:sns}
\end{figure}

Concerning the interface between the superconducting and the normal
regions, two scenarios have to be distinguished: the case of weak
coupling, realized for instance by tunneling barriers\cite{brouwer97}
and the case of strong coupling with ideal, impurity-free leads at
$z_1$ and $z_2$ (see Fig.\ \ref{fig:sns}). 
%In the first case, the
%distribution of the random scattering matrix for the junction is
%characterized by a Poisson kernel.\cite{beenakker97,krieger} 
In the latter case, the scattering matrix $\mathcal{S}$ is a unitary
random matrix with eigenvalues uniformly distributed on the complex
unit circle,\cite{beenakker97,baranger} and the corresponding ensemble
is referred to as CUE. In this article we will consider only the
circular ensemble, that is, we stay in the strong coupling regime.
Furthermore, we assume the Thouless energy to be much larger than the
gap, $E_c \gg \Delta$. In this limit, which was named ``short
dwell-time regime'' in Ref.~\onlinecite{brouwer97}, the energy
dependence of the scattering matrix can be neglected.

%%%% Diesen Abschnitt wuerde ich am liebsten weglassen %%%%
On the other hand, the dwell-time $\tau_d$ of electrons and holes in
the chaotic region --- with energies close to the Fermi level -- has
to be sufficiently large so that RMT is applicable. The dwell-time
$\tau_d$ should be larger than the time scale $\tau_{\rm erg}$ at
which ergodicity sets in, where $\hbar/\tau_{\rm erg} = \hbar
v_{\mbox{\tiny{F}}} \mbox{min}\{l,L\}/L^2$.
%%%%%%%%%%%%%%%%%%%%%%%%%%%%%%%

The scattering matrix $\mathcal{S}$ has the block structure
\begin{equation}
\mathcal{S} = \left( \begin{array}{cc} r & t' \\ t & r' \end{array} \right) \,,
\end{equation}
with the reflection matrices $r,\ r'$ and the transmission matrices
$t,\ t'$, all of which are $N \times N$ matrices. The number of
propagating modes in each of the two leads is $N$. Current
conservation implies that the scattering matrix is unitary:
$\mathcal{S}^{-1} = \mathcal{S}^{\dagger}$, and a ``polar
decomposition'' is possible:\cite{mello,martin}
\begin{equation}\label{polDekomp}
\mathcal{S} = 
\left( \begin{array}{cc} U & 0 \\ 0 & V \end{array}\right) 
\left( \begin{array}{cc} -\sqrt{1-T} & \sqrt{T} \\ 
                        \sqrt{T} & \sqrt{1-T} \end{array}\right) 
 \left( \begin{array}{cc} U' & 0 \\ 0 & V' \end{array}\right)\,,
\end{equation}
where a $N \times N$ diagonal matrix $T =$ diag($T_1,T_2,\ldots,T_N$)
has been introduced. The transmission eigenvalues $T_n$ are the
eigenvalues of the matrix product $tt^{\dagger}=t't^{'\dagger}$. The
four unitary $N \times N$ matrices $U,U'$ and $V,V'$ are to be
interpreted as ``phase factors''. In the case of time-reversal symmetry the
scattering matrix is unitary and symmetric, $\mathcal{S} =
\mathcal{S}^{T}$, which implies that also $U' = U^{T}$ and $V' = V^{T}$
hold (``circular orthogonal ensemble'' (COE)).

For statistical averaging we need the Haar measure $\mu$ which can be
determined for the above polar decomposition
as\cite{baranger,jalabert}
\begin{equation}\label{HaarMeasure}
 d\mu(\mathcal{S}) = J(\{T_n\}) \prod_{\alpha} d\mu(U_{\alpha}) \prod_i d T_i\,. 
\end{equation}
The Jacobian is the probability distribution $P(\{T_n\})$ of the transmission
eigenvalues (up to a normalization constant $c_N$); it is given by
\begin{equation} \label{Tprob}
J(\{T_n\})= \prod_{n<m} |T_n - T_m |^{\beta} \prod_k T^{-1+\beta/2}_k = c_N\, P(\{T_n\})
\end{equation}
Here, $\beta$ is the ensemble parameter---which numbers the degrees of
freedom of the matrix elements of $\mathcal{S}$:
\begin{eqnarray}
\beta &=& 1 \qquad \mbox{COE}\qquad \mbox{system with time-reversal symmetry}\nonumber\\
\beta &=& 2 \qquad \mbox{CUE}\qquad \mbox{no time-reversal symmetry} 
\end{eqnarray}
for the two cases considered in this paper. The set of
independent unitary matrices in the Haar measure (\ref{HaarMeasure})
is $\{U_{\alpha}\}=\{U,V,U',V'\}$ for $\beta=2$, and
$\{U_{\alpha}\}=\{U,V\}$ for $\beta=1$.  For the purpose of
interpreting the probability distribution (\ref{Tprob}) it has become
popular to write it in the form of a Gibbs distribution
\begin{equation}
P(\{T_n\})=\; {1\over c_N}\;
     \exp\left(\beta \sum_{n<m} \ln |T_n - T_m| - \beta \sum_k V(T_k) \right)\,,
\end{equation}
with the ``one-particle'' potential $V(T_k) = (1/\beta - 1/2) \ln T_k$ and
a ``two-particle'' logarithmic repulsion.

\subsection{Scattering Matrix of the SNS Junction}
Here we follow the approach pioneered by Beenakker,\cite{beenakker92}
and extend it in order to include the case of heterogeneous
junctions.\cite{comment1} The details of how to set up the Bogoliubov-de Gennes (BdG)
equation for this type of junction are found in the cited paper and in
Ref.~\onlinecite{boettcher}, and will not be repeated here.

The S-matrix of the complete junction,
$\mathbf{S}_{\mbox{\tiny{SNS}}}$, connects in- with out-going
amplitudes of the electronic wave function in the superconductors.
The components of the wave function are defined in Fig.~\ref{fig:sns};
for example, we have for the outgoing amplitudes $N$ components
$a^-_e$ for electron-like quasi-particles running from the interface
$z_1$ to the left and $N$ components $d^+_e$ for electron-like
quasiparticles running from the interface $z_2$ to the right. Andreev
reflection mixes particle with hole components, and therefore the corresponding
hole-like quasiparticles with $N$ components $a^+_h$ and $d^-_h$ have
to be included in the scattering process. The $4N\times 4N$-matrix
$\mathbf{S}_{\mbox{\tiny{SNS}}}$ is now defined through
\begin{equation}
\left( \begin{array}{c} a^-_e \\ d^+_e \\ a^+_h \\ d^-_h 
\end{array} \right) = \mathbf{S}_{\mbox{\tiny{SNS}}}
\left( \begin{array}{c} a^+_e \\ d^-_e \\ a^-_h \\ d^+_h 
\end{array} \right)\,.
\end{equation}
The matrix $\mathbf{S}_{\mbox{\tiny{SNS}}}$ is actually constructed
from more basic $4N\times 4N$ S-matrices,
$\mathbf{S}_{\mbox{\tiny{N}}}$ and $\mathbf{S}_{\mbox{\tiny{A}}}$.
The matrix $\mathbf{S}_{\mbox{\tiny{N}}}$ is formed by the scattering
matrices for electrons and holes in the normal, disordered region:
\begin{equation}\label{MatrixDisorder}
\mathbf{S}_{\mbox{\tiny{N}}}(E) = \left( \begin{array}{cc} \mathcal{S}(E) & 0 \\
0 & \mathcal{S}^*(-E) \end{array} \right)
\end{equation}
where $E$ is the energy relative to the Fermi energy.
The matrix $\mathbf{S}_{\mbox{\tiny{A}}}$, defined by 
\begin{equation}\label{MatrixAndreev}
\mathbf{S}_{\mbox{\tiny{A}}}(E) = 
\left( \begin{array}{cc} 0 & e^{-i (\mathbf{\varphi}(E) - \mathbf{\Phi})} \\ 
e^{-i (\mathbf{\varphi}(E) + \mathbf{\Phi})} & 0 \end{array} \right)
\end{equation} 
is the scattering matrix for Andreev reflections at both NS
interfaces.  The off-diagonal structure originates in the fact that
Andreev reflection transforms an electron into a hole. The
transformation is supplemented by two phase shifts: one from the
penetration of the wave function into the superconductors
\begin{equation}
\mathbf{\varphi} = \left( \begin{array}{cc} \varphi_1\,\mathbf{1}_N & 0 \\
0 & \varphi_2\,\mathbf{1}_N \end{array} \right) 
\end{equation}
where  $\mathbf{1}_N$ is the $N \times N$ unit matrix, and 
\begin{equation}\label{PhasePhi}
\varphi_j = -i\,\ln\left(\frac{E}{\Delta_j} + \sqrt{\left(\frac{E}{\Delta_j}\right)^2 -1}\right)\,.
\end{equation}
The $\Delta_j$ are the respective Bogoliubov quasiparticle energy gaps
of the two superconductors. We use rigid boundary conditions with
step-like gaps at the SN-interfaces. This is approximately valid for
the considered case that the extension of the junction is much smaller
than the coherence length in the bulk superconductor.\cite{likharev}
The second contribution to the phase shift in (\ref{MatrixAndreev}) is
the phase $\phi_{1,2}$ acquired from the pair potential in the
superconductors which has the matrix form
\begin{equation}\label{MatrixPhi} 
\mathbf{\Phi} = \left( \begin{array}{cc} \phi_1\,\mathbf{1}_N & 0 \\
0 & \phi_2\,\mathbf{1}_N \end{array} \right)\,.
\end{equation} 
With these basic elements the scattering matrix for multiple Andreev
reflection at the interfaces, in combination with the intermediate
propagation through the disordered region, takes the form\cite{beenakker92}
\begin{eqnarray}\label{StreuSNS}
\mathbf{S}_{\mbox{\tiny{SNS}}} & = &
\left( \begin{array}{cc}e^{i(\mathbf{\varphi}-\mathbf{\Phi})/2} & 0 \\ 
0 & e^{i(\mathbf{\varphi}+\mathbf{\Phi})/2} \end{array} \right) 
\mathbf{S}_{\mbox{\tiny{N}}}(E)
[ \mathbf{1}_{4N} - \mathbf{S}^\dagger_{\mbox{\tiny{N}}}(E) 
\mathbf{S}_{\mbox{\tiny{A}}}(E) ] \nonumber\\
& & [ \mathbf{1}_{4N} - \mathbf{S}_{\mbox{\tiny{A}}}(E)
\mathbf{S}_{\mbox{\tiny{N}}}(E) ]^{-1}
\left( \begin{array}{cc}e^{-i(\mathbf{\varphi}-\mathbf{\Phi})/2} & 0 \\ 
0 & e^{-i(\mathbf{\varphi}+\mathbf{\Phi})/2} \end{array} \right)
\end{eqnarray}
where $\mathbf{1}_{4N}$ is the $4N \times 4N$ unit matrix.  Obviously,
the $\varphi_j$ defined in (\ref{PhasePhi}) are real for $E <
\min\{\Delta_1,\Delta_2\}\equiv {\Delta_{\mbox{\tiny{min}}}}$.  In
this case, the matrix $\mathbf{S}_{\mbox{\tiny{A}}}$ is unitary:
Andreev reflection within the region between the two interfaces is
current conserving. For $|E| > \max\{\Delta_1,\Delta_2\} \equiv
{\Delta_{\mbox{\tiny{max}}}}$, the matrix
$\mathbf{S}_{\mbox{\tiny{A}}}$ is hermitian, and only then is
$\mathbf{S}_{\mbox{\tiny{SNS}}}$ a unitary matrix.

\subsection{Spectrum of States of a SNS-Junction}
Corresponding to the range of energies at which we consider the
electron propagation through the junction, we encounter three
different regimes:
\begin{enumerate}
\item $|E| > {\Delta_{\mbox{\tiny{max}}}}$, in which case we consider a
contiuum of particle and hole scattering states in each
superconductor,

\item $ {\Delta_{\mbox{\tiny{max}}}} > E >
{\Delta_{\mbox{\tiny{min}}}}$, which is the intermediate regime with
quasiparticle states extending over the superconductor with the lower
energy gap but they decay exponentially in the superconductor with the
larger gap,

\item $|E| < {\Delta_{\mbox{\tiny{min}}}}$, in which case the BdG wave
functions decay exponentially in both superconductors. Bound states,
{\it Andreev states}, which are localized in the normal region,
generate a discrete spectrum. For a finite phase difference, $\phi =
\phi_2 - \phi_1$, Andreev states support an electrical current.
Thereby, the coherent particle-hole excitations in the junction are
converted into a supercurrent in the bulk superconductors.
\end{enumerate}

\subsubsection{Continuum states}
In order to obtain the spectrum in this regime, one makes use of the
``Wigner time-delay matrix''\cite{goldberger64,doron92}
$\mathbf{S}^\dagger \frac{\partial}{\partial E} \mathbf{S}$, the trace
of which is the density of states, $\rho(E)$:
\begin{equation}\label{WignerDensity}
\rho(E) = \frac{1}{2\pi i} \mbox{tr} \left\{ \mathbf{S}^\dagger
\frac{\partial}{\partial E} \mathbf{S}  \right\} = 
\frac{1}{2\pi i}\frac{\partial}{\partial E}\ln\det \mathbf{S}\,.
\end{equation}
As the scattering matrix of the junction
$\mathbf{S}_{\mbox{\tiny{SNS}}}$ is unitary, one may evaluate the
relation (\ref{WignerDensity}) with $\mathbf{S}
=\mathbf{S}_{\mbox{\tiny{SNS}}}$ and find, respecting the hermiticity
of $\mathbf{S}_{\mbox{\tiny{A}}}(E)$,
\begin{equation}\label{DOS}
\rho(E)  =  - \frac{1}{\pi}\frac{\partial}{\partial E} \Im \ln Z(E) + \rho_0(E)
\end{equation}
with $Z(E)$ a complex function defined through
\begin{equation}\label{functionZ}
Z(E) = \mbox{det}( \mathbf{1}_{4N} - \mathbf{S}_{\mbox{\tiny{A}}}(E)
\mathbf{S}_{\mbox{\tiny{N}}}(E))\,.
\end{equation}
The additive term 
\begin{equation}
\rho_0(E) = \frac{1}{2\pi i}\frac{\partial}{\partial E}\ln\det \mathbf{S}_{\mbox{\tiny{N}}}
\end{equation} 
in the density of states (\ref{DOS}) is independent of the phase
difference $\phi$ of the two superconductors and will not contribute
to the Josephson current.

\subsubsection{Intermediate regime}
The determination of the density of states in this energy regime does
not proceed parallel to that of $\rho(E)$ with extended
quasiparticle states in both superconductors.  The scattering matrix
of the junction is not unitary for
${\Delta_{\mbox{\tiny{max}}}}>\,E>\,{\Delta_{\mbox{\tiny{min}}}}$ since
the quasiparticle states decay exponentially in the superconductor
with the larger energy gap --- we assume this to be the left
superconductor in Fig.~\ref{fig:sns}. However, we can define a reduced
scattering matrix $\mathbf{S}_{\mbox{\tiny{int}}}$ which connects only
the incoming amplitudes from the right superconductor to the outgoing
amplitudes of the right superconductor
\begin{equation}
{d^+_e \choose d^-_h} = \mathbf{S}_{\mbox{\tiny{int}}} {d^-_e \choose d^+_h}\,,
\end{equation}
and the dimension of this scattering matrix is half that of
$\mathbf{S}_{\mbox{\tiny{SNS}}}$.  Since
$\mathbf{S}_{\mbox{\tiny{int}}}$ is unitary, we may use relation
(\ref{WignerDensity}) for $\mathbf{S}
=\mathbf{S}_{\mbox{\tiny{int}}}$. After some tedious algebra we arrive
again at
\begin{equation}\label{DOS-int}
\rho(E)  =  - \frac{1}{\pi}\frac{\partial}{\partial E} \Im \ln Z(E) + \tilde{\rho}_0(E)
\end{equation}
with a $\phi$-independent additive term $\tilde{\rho}_0$
\begin{equation}
\tilde{\rho}_0(E) = \frac{1}{2\pi i}\frac{\partial}{\partial E}\ln\det\mathbf{S}_{\mbox{\tiny{N}}}
  - \, {\varphi_1(E)\over\pi}\,.
\end{equation} 
As already indicated, the scattering matrix for Andreev scattering
$\mathbf{S}_{\mbox{\tiny{A}}}(E)$ in $Z(E)$ is hermitian for the
continuum states but not so for the intermediate states, and
expressions (\ref{DOS}) and (\ref{DOS-int}) are indeed quite distinct.

\subsubsection{Bound states}
Since the Andreev states are evanescent in the superconductors, the
matrix $\mathbf{S}_{\mbox{\tiny{A}}}(E)$ is now unitary. It obeys the
simple relation for bound states
\begin{equation}
\left( \begin{array}{c} b^+_e \\ c^-_e \\ b^-_h \\ c^+_h \end{array} \right)\ =
\mathbf{S}_{\mbox{\tiny{A}}}(E) \mathbf{S}_{\mbox{\tiny{N}}}(E)
\left( \begin{array}{c} b^+_e \\ c^-_e \\ b^-_h \\ c^+_h \end{array} \right)\ 
\end{equation}
Solutions exist if $Z(E)=0$ is fulfilled.
In this regime the density of states can therefore be written as
\begin{eqnarray}
\rho(E) &=& %\sum_i \delta(E-E_i) = 
\left|\frac{\partial Z}{\partial E} \right| \delta(Z(E))
= - \frac{1}{\pi}\, \frac{\partial Z}{\partial E}\, \Im \frac{1}{Z(E + i\,0)}  \nonumber \\
&=&  - \frac{1}{\pi}\,\frac{\partial}{\partial E}\, \Im \ln Z(E + i\,0) 
\end{eqnarray}

\subsection{Josephson Current}
The stationary current $I$ through a SNS-junction is driven by the
phase difference of the superconductors. It can be obtained from a
free energy $F$ through the well-known thermodynamic relation
\begin{equation}\label{josephson-current}
I = \frac{2e}{\hbar}\,\frac{dF}{d\phi}\,.     
\end{equation}
As elaborated in the literature (for example, see
Ref.~(\onlinecite{beenakker92})) the current is given in terms of the
density of quasiparticle states
\begin{equation}
I = - \frac{2e}{\hbar}\; 2 k_B \tau\, \int\limits_{0}^{\infty} dE 
\ln{\left[2 \cosh{\left(E/ 2 k_B \tau\right)}\right]}\; \frac{d}{d \phi} \rho(E)\,.
\end{equation}
After an integration by parts the current can be expressed as 
\begin{equation}\label{super-current}
I = \frac{2e}{\pi \hbar}\; \int\limits_{0}^{\infty} dE \tanh{\left[E/ 2 k_B \tau\right]}\, j(E+i\,0,\phi)\,,
\end{equation}
where $j$ is a kind of spectral current
\begin{eqnarray}\label{spectral-current}
j(E,\phi) &=& -\frac{d}{d \phi} \Im \ln Z(E) \nonumber\\
&=&  -\frac{d}{d \phi} \Im \ln \mbox{det}( \mathbf{1}_{4N} - 
\mathbf{S}_{\mbox{\tiny{A}}}(E)\mathbf{S}_{\mbox{\tiny{N}}}(E))\,.
\end{eqnarray}
This compact expression for the supercurrent is the basis for the
calculation of the distribution of supercurrents through chaotic
junctions in the following sections.

\section{SINGLE CHANNEL CASE}
Quantum transport of coherent particle-hole excitations through a
single channel is rather particular. Level repulsion is missing by
construction. The transmission eigenvalues $T$ are distributed
according to a ``single-particle'' potential which is either
logarithmically diverging for small $T$ ($\beta=1$) or constant
($\beta=2$), $P(T) = \frac{\beta}{2}\, T^{-1+\beta/2}$.  This
potential will control the distribution of the critical currents. It
should be realized that this scenario applies also for junction
geometries where (many) different channels are uncoupled.

It is quite instructive to derive step by step the spectral current
for this single channel case since the procedure conveys an
understanding of the validity and consequences of the approximations
made. Here we present the result for the case of a scattering matrix
which is energy independent, that is, the Thouless energy is much larger
than $\Delta_{\mbox{\tiny{max}}}$, and there are no resonances at or
in the gap. The following relation holds
\begin{equation}\label{GreenF-zeroE}
j(E,\phi) = - \frac{d}{d\phi}\Im\ln [\cos(\varphi_1(E)\!+\varphi_2(E))
\!- R \cos(\varphi_1(E)\!-\varphi_2(E))\!- T \cos\phi_{\rm eff} ], \nonumber
\end{equation}
which is derived in Appendix~A where we also discuss an expansion in
$E/E_{\mbox{\tiny{F}}}$ of the phases of the scattering matrix. In the above
expression, we introduced $R=1-T$ and $\phi_{\rm eff}=\phi +
\Delta\phi$, where $\Delta\phi$ is a phase which is generated from the
phase factors $U,U'$ and $V,V'$ in the polar decompostion
(\ref{polDekomp}). Since $\Delta\phi$ builds only on a difference of the
respective phases, it is zero for the COE where $U' = U^{T}$ and $V' =
V^{T}$ hold, whereas it runs from $0$ to $2\pi$ for the CUE.

We now restrict our discussion to symmetric contacts with
$\Delta_1=\Delta_2=\Delta$.  The energy eigenvalues for the Andreev
states are straightforwardly found from the zeros of the logarithm in
(\ref{GreenF-zeroE}),
%\begin{equation}\label{Andreev-energies},
$
E(\phi) = \pm\Delta \sqrt{1-T\sin^2\frac{\phi}{2}}
$.
%\end{equation} 
The supercurrent (\ref{super-current}) is carried only by the Andreev
states (the continuum contributions cancel):\cite{haberkorn78}
\begin{equation}\label{OneChannelCurrent}
I(\phi,T) = \frac{e \Delta}{2 \hbar} \frac{T \sin\phi}{\sqrt{1-T\sin^2\frac{\phi}{2}}}  
\tanh(\frac{\Delta}{2 k_B \tau}\sqrt{1-T\sin^2\frac{\phi}{2}})\,.
\end{equation}
Obviously, this expression recovers the well known sine-phi dependence
for small transmission coefficients, $I(\phi) = \frac{\pi\Delta}{2 e}
G \sin\phi$, with the conductance $G = 2 e^2 T/h$.

Finally, the current for a particular junction
(\ref{OneChannelCurrent}), characterized by a specific $T$, is
averaged with the distribution function (\ref{Tprob}) for a single
channel and we find for
$\tau=0$ and $\beta=1$ (see Fig.~\ref{fig:iav-rms}, lowest curve)
\begin{equation} \label{OneChannelCurrentAv} 
\langle I(\phi) \rangle_{\mbox{\tiny{COE}}} =\frac{e \Delta}{\hbar} \sin\phi\, 
\frac{\phi-\sin\phi}{8 \sin^3\frac{\phi}{2}}
\qquad\mbox{for $\phi \in [-\pi,\pi]$}
\end{equation}
For $\beta=2$, the phase difference $\phi$ in
(\ref{OneChannelCurrent}) has to be replaced by $\phi_{\rm eff}$.
Integration over $d\mu(U_{\alpha})$ (see Eq.~(\ref{HaarMeasure}))
corresponds to averaging over $\Delta\phi$ which results in a zero
ensemble-averaged supercurrent for the CUE.

Moreover, the variance of the supercurrent can be calculated analytically.
However we want to focus on a different issue: the critical current,
that is, the maximum supercurrent is usually more accessible in
experiments. In the single-channel case the structure of the
supercurrent-phase relation (\ref{OneChannelCurrent}) is actually so
elementary for zero temperature that it is possible to determine
directly the distribution function for the critical current.  The
distribution function $\rho(i_c)$ for the dimensionless critical
current, $i_c = I_c / \frac{e \Delta}{\hbar}$, characterizes the
ensemble in a unique way, as we will present below.
\begin{table}
\setlength{\tabcolsep}{1cm}
\renewcommand{\arraystretch}{2.5}
\begin{center}
\begin{tabular}{r|c|c}
                     & CUE   & COE \\ \hline
$P(T)$                 & $1$            & $\displaystyle{ \frac{1}{2 \sqrt{T}}}$ \\
$\rho(i_c)$            & $2\,(1-i_c)$   & $\displaystyle{\frac{(1-i_c)}{\sqrt{i_c(2-i_c)}}}$ \\
$\langle i_c \rangle$  & $\displaystyle{\frac{1}{3}}$  & $\displaystyle{1-\frac{\pi}{4}}$ \\
$\mbox{var}\,i_c$     & $\displaystyle{\frac{1}{18}}$ & $\displaystyle{\frac{2}{3} - \frac{\pi^2}{16}}$ \\
\end{tabular}
\end{center}
\bfseries   
\caption{\mdseries\em\label{tab:Disbrib-Ic} 
Disbribution of the critical currents for the CUE und COE; the average critical current is
$\langle i_c \rangle= \int_0^1 i_c\, \rho(i_c) $, and the variance
is defined as $\mbox{\rm var}\,i_c = \langle i_c{}^2 \rangle - \langle i_c \rangle^2$.
}
\end{table}
A suitable definition of the distribution function for the critical
current is
\begin{equation}\label{Def-rho}
\rho(i_c) = \int\limits^1_0 dT\,P(T)\,\delta(i(T,\phi_c(T))-i_c)
\,.
\end{equation} 
The critical current for a specific transmission coefficient,
$i(T,\phi_c)$, is found from (\ref{OneChannelCurrent}) (at $\tau=0$)
together with the condition $d\,i(\phi_c,T) /d\phi_c = 0$. It yields
an implicit equation for $\phi_c$ which reads $T=T_c(\phi_c) = -
\cos\phi_c/\sin^4\frac{\phi_c}{2}$. This latter relation allows to
express the critical current in terms of $\phi_c$: it reads
$i(T_c(\phi_c),\phi_c) = 1 - (\tan\frac{\phi_c}{2})^{-2}$.  Since the
function $T_c(\phi_c)$ is monotonic and maps the interval
$[\pi/2,\pi]$ onto the interval $[0,1]$ we may rewrite the
integral~(\ref{Def-rho}) in terms of an integration over $\phi_c$.
Finally, making use of $T_c(\phi_c) = i_c (2-i_c)$, we find the
fundamental zero temperature relation for the distribution of the
critical current:
\begin{equation} \label{Distrib-Ic}
\rho(i_c) = 2\,(1-i_c)\,P(i_c\,(2-i_c))\,.
\end{equation}
Again, $i_c$ is normalized to the maximum physical critical current
${e \Delta}/{\hbar}$, so it covers the interval $[0,1]$. We would like
to emphasize that this formula is valid in the presence as well as in
the absence of time-reversal symmetry. For the single-channel case
breaking of time-reversal symmetry results in a shift of the $I(\phi)$
curve by a constant phase $\Delta\phi$ which does not affect the
critical current.

The immediate consequences for the COE and CUE are summarized in
Table~\ref{tab:Disbrib-Ic}.  Relation~(\ref{Distrib-Ic}) allows to
determine the type of ensemble if the distribution of critical
currents is measured. The distribution function for the COE diverges as $1/\sqrt{i_c}$
for $i_c\rightarrow 0$ (which is also displayed in
Fig.~\ref{fig:rhoic}, $N=1$), whose origin is the square root
divergence of the distribution function $P(T)$.

\section{MULTIPLE CHANNEL JUNCTIONS}
Only in the presence of time-reversal symmetry, $\beta = 1$, the
expression (\ref{GreenF-zeroE}) for the spectral current can be
generalized to multiple channels,
\begin{equation}\label{MultiChannelSpectCurrent}
j(E,\phi)\!=\!-\!\sum\limits^{N}_{n=1}\frac{d}{d\phi}\Im\ln [\cos(\varphi_1(E)\!+\varphi_2(E))
\!-R_n \cos(\varphi_1(E)\!-\varphi_2(E))\!-T_n \cos\phi ]. \nonumber
\end{equation}
In Appendix B the averaged critical supercurrent for the asymmetric
Josephson junction is computed from this expression.  For the
symmetric junction, $\Delta_1 = \Delta_2 = \Delta$, again only the
Andreev-states contribute and the current is given as a linear
statistic as obtained by Beenakker,\cite{beenakker92}
\begin{equation}\label{MultiChannelCurrent}
I(\phi,\{ T_i \}) = \frac{e \Delta}{\hbar} \sum\limits^{N}_{n=1} 
\frac{T_n \sin\phi}{2 \sqrt{1-T_n\sin^2\frac{\phi}{2}}}  
\tanh(\frac{\Delta}{2 k_B \tau}\sqrt{1-T_n\sin^2\frac{\phi}{2}})\,.
\end{equation}

The average supercurrent can be determined analytically in the limit
of large channel number $N$ and zero temperature.\cite{baranger} It is
obtained with the density of transmission coefficients, $\rho(T) =
N/\pi\,[T(1-T)]^{-1}$, and is non-sinosoidal due to its dependence on
the hypergeometric function $_2F_1$,
\begin{equation}
\langle I(\phi) \rangle_{\mbox{\tiny{COE}}} = N\,\frac{e \Delta}{4 \hbar}\, 
_2F_1(\frac{1}{2},\frac{3}{2},2,\sin^2\frac{\phi}{2})\, \sin\phi \quad \mbox{for} \quad N \rightarrow \infty.
\end{equation}

\begin{figure}
\vspace*{1cm}
\begin{center}
   \psfrag{N=inf}{\tiny ${}\!\!N\!\!=\!\infty$}
   \psfrag{N=1}{\tiny ${}\!N\!\!=\!1$}
   \psfrag{N=2}{\tiny ${}\!N\!\!=\!2$} 
   \psfrag{N=3}{\tiny ${}\!N\!\!=\!3$} 
   \psfrag{N=4}{\tiny ${}\!N\!\!=\!4$} 
   \psfrag{N=5}{\tiny ${}\!N\!\!=\!5$}
   \psfrag{1}{\small $0$} 
   \psfrag{1.5}{\small $\pi/2$} 
   \psfrag{2}{\small $\pi$}  
   \psfrag{0}{\small $0$}
   \psfrag{0.1}{\small $0.1$}
   \psfrag{0.2}{\small $0.2$}
   \psfrag{0.3}{\small $0.3$}
   \psfrag{0.4}{\small $0.4$}
   \psfrag{supercurrent}{\small $\langle I \rangle$ [$Ne\Delta/\hbar$]}
   \psfrag{fluctuation}{\small rms $I$ [$e\Delta/\hbar$]}
   \psfrag{p}{\small $\phi$}

   \psfrag{(a)}{(a)}
   \psfrag{(b)}{(b)}
     \leavevmode

\includegraphics[height=2.1in]{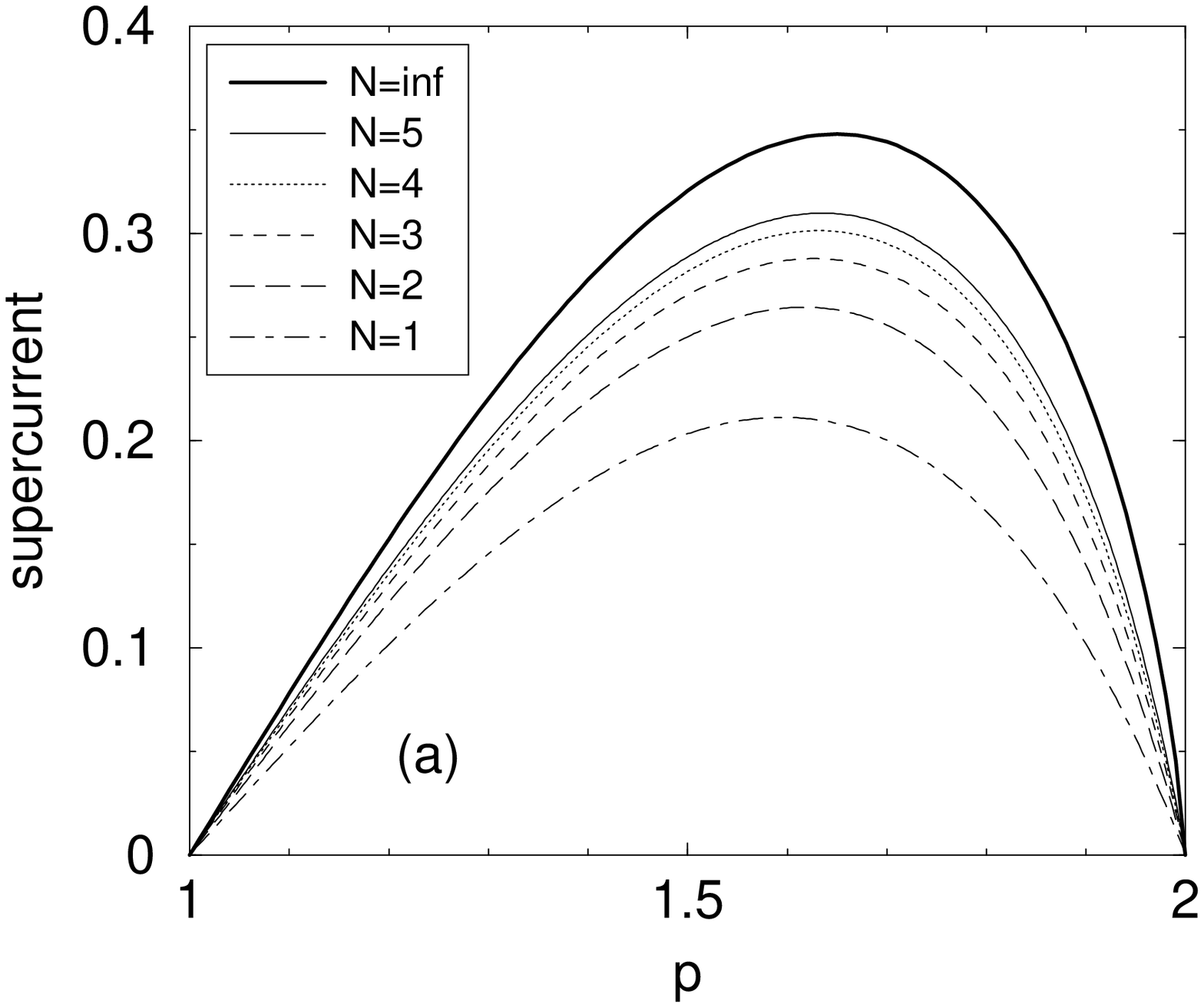}
\includegraphics[height=2.1in]{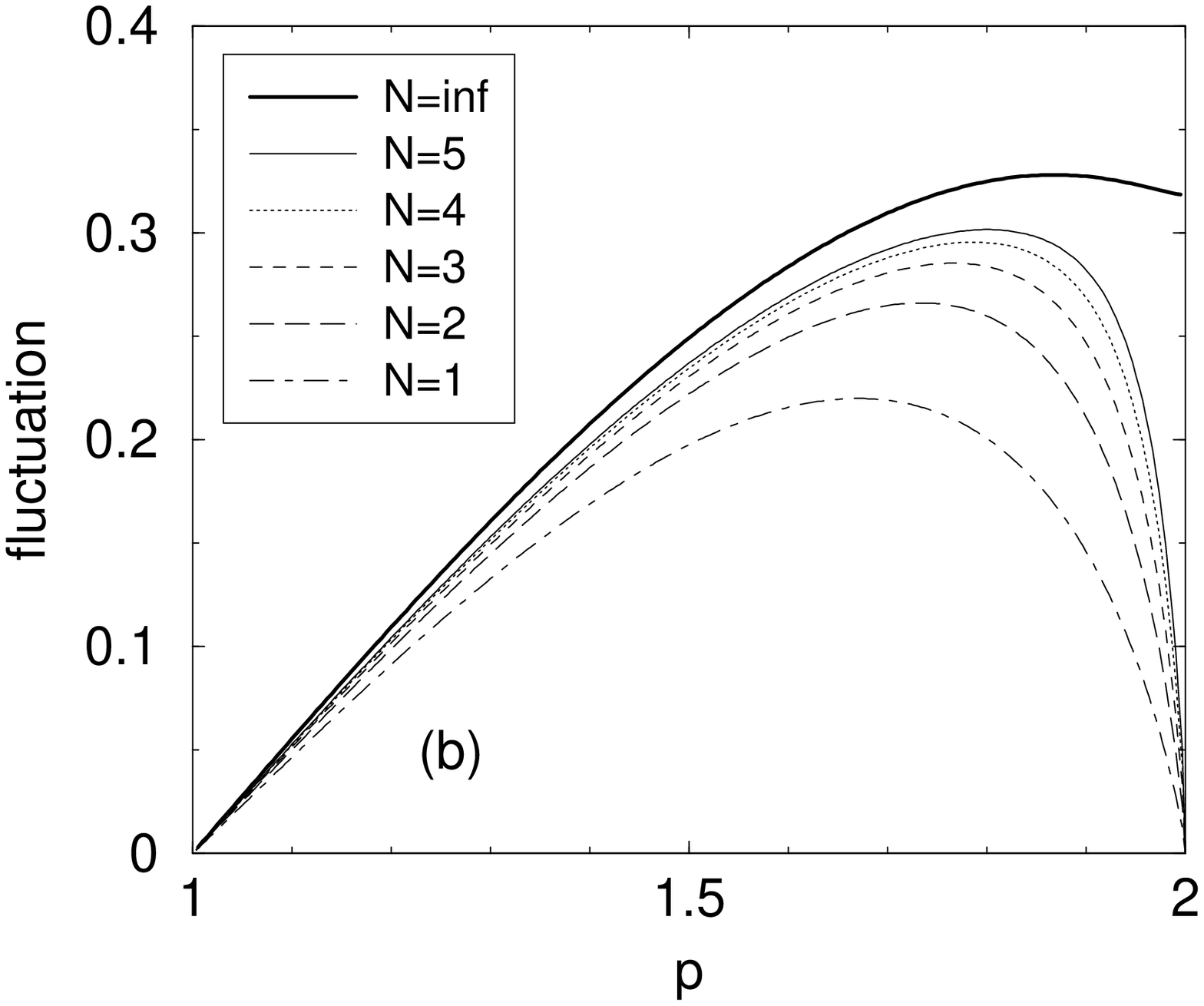}
\end{center}
\caption{Average supercurrent for the COE, scaled with $Ne\Delta/\hbar$, as a function of
  the phase difference $\phi$ across the junction (a), and the rms
  fluctuations of the supercurrent, also for the COE (b).}
\label{fig:iav-rms}
\end{figure}
This expression, valid for large $N$, is to be compared to results for
finite $N$. For this purpose, the average supercurrent has been
evaluated numerically for $N=1$ to $N=5$. Although the curves do not
scale with $N$---the maxima are slightly shifted with increasing $N$
towards higher values of $\phi$ (see Fig.~\ref{fig:iav-rms}a)---the
overall shape of the supercurrent-phase relation is remarkably similar
for $N=1$ to $N=\infty$. The same holds, to a lesser extent, for the
root-mean-square (rms) fluctuations (Fig.~\ref{fig:iav-rms}b). For
$N=\infty$, we included the result obtained by
Beenakker,\cite{beenakker93} which is slowly approached by the
finite-$N$ results.  He argued that corrections in $1/N$ should move
the finite value of rms$\ I_{N=\infty}(\phi=\pi)=e\Delta/\pi\hbar$ to zero.

Much more pronounced is the channel-number dependence of the
distribution function for the critical current. The divergence of the
distribution function in the COE is lifted in the multiple-channel
situation due to the interaction between the transmission eigenvalues,
that is, due to level repulsion. It is not possible to derive a closed
expression for the distribution in the case of two or more channels.
However, an analytical expression can be found which controls the
limit $i_c\rightarrow0$. The proof exploits again the simplicity of
the current-phase relation (\ref{MultiChannelCurrent}) where we
restrict the evaluation to zero temperature. The limit
$i_c\rightarrow0$ is approached when all $T_n\rightarrow0$.  In this
case, we neglect the square-root denominator, which results in
$\phi_c\rightarrow\pi/2$. As expected, the critical current is in this
limit proportional to the conductance $g$, i.e. the sum over all
transmission eigenvalues,
$i(\phi_c,\{T_i\})=\frac{1}{2}\sum\nolimits^{N}_{n=1} T_n =
\frac{1}{2} g$. Therefore, for small $i_c$, the distribution of the 
critical current behaves in the same manner as the distribution of
the conductance does.\cite{baranger}

Similar to the single-channel situation, the distribution function for
the critical current is defined as
\begin{equation} \label{Def-multi-rho}
\rho(i_c)\; =\;  c^{-1}_N\; \prod_l \int_0^1 dT_l \prod_{n<m} |T_n - T_m | 
\prod_k T^{-1/2}_k \;\delta\,\bigl(\,i(\phi_c(\{T_i\}),\{T_i\})-i_c\bigr)
\end{equation}
After rescaling all transmission coefficients in (\ref{Def-multi-rho}) with
$i_c$ one obtains the limiting small-current behavior
\begin{equation} \label{Def-limit-rho}
\rho(i_c\rightarrow0)\; \sim\; i_c{}^\alpha\qquad{\rm with}\qquad
% \alpha\;=\;\biggl[\frac{N}{2}+{N\choose2}\biggr] -1
 \alpha\;=\;\frac{N^2}{2}-1
\end{equation}
For a single channel, $N=1$, this agrees with the result of
Table~\ref{tab:Disbrib-Ic}. In particular for two channels one obtains
\begin{equation}
\rho(i_c\rightarrow0)\sim i_c \qquad{\rm for }\qquad N=2\,.
\end{equation}
Level repulsion is responsible for the vanishing distribution function
in the limit of zero critical current.

\begin{figure}
%\vspace*{1cm}
%\begin{center}
   \psfrag{N=1}{\small N=1}
   \psfrag{N=2}{\small N=2} 
   \psfrag{N=3}{\small N=3} 
   \psfrag{N=4}{\small N=4} 
   \psfrag{N=5}{\small N=5} 
   \psfrag{N=10}{\small N=10}

   \psfrag{0}{\small $0$}
   \psfrag{0.2}{\small $0.2$}
   \psfrag{0.4}{\small $0.4$}
   \psfrag{0.6}{\small $0.6$}
   \psfrag{0.8}{\small $0.8$}
   \psfrag{1}{\small $1$}
   \psfrag{5}{\small $5$}
   \psfrag{10}{\small $10$}
   \psfrag{15}{\small $15$}
   \psfrag{0.5}{\small $0.5$}
   \psfrag{1.5}{\small $1.5$}
   \psfrag{2}{\small $2$}
   \psfrag{3}{\small $3$}
   \psfrag{-1}{\small $-1$}
   \psfrag{-2}{\small $-2$} 
   \psfrag{20}{\small $20$}
   \psfrag{30}{\small $30$}
   \psfrag{-10}{\small $-10$}
   \psfrag{-20}{\small $-20$}

   \psfrag{critical current}{\small $I_c\,\, [Ne \Delta/\hbar]$} 
   \psfrag{critical c2}{\small $I_c\,\, [e \Delta/\hbar]$}
   \psfrag{d}{\small $\rho\, (I_c)$}
   \psfrag{d2}{\small $\rho$} 
   \psfrag{dd}{\small $\rho'$}
   
     \leavevmode
\includegraphics[width=4.7in,height=3.3in]{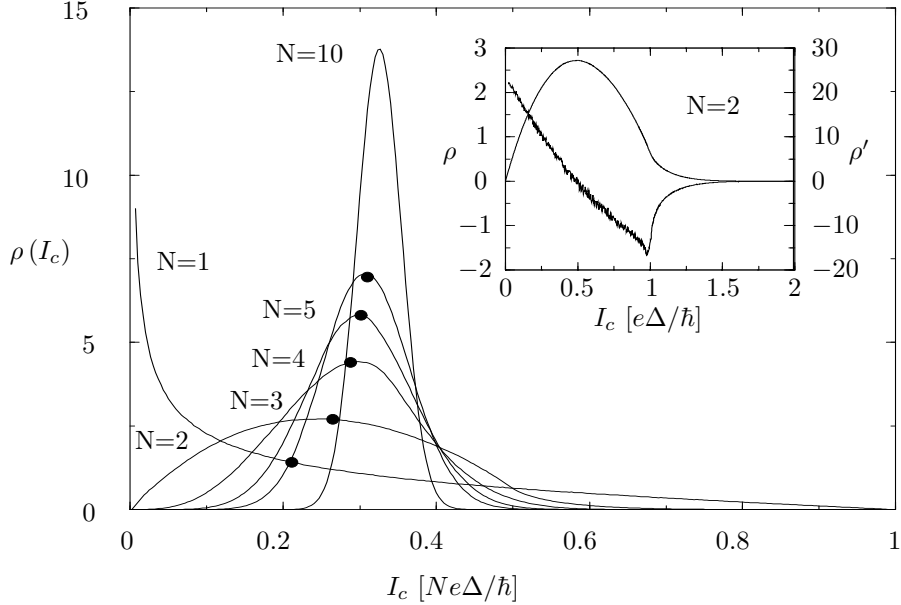}
%\end{center}
\caption{Distribution function for the critical current in the COE for various 
  channel numbers $N$. The maximum value reached by the averaged
  $\langle I(\phi) \rangle$ (Fig.~\ref{fig:iav-rms}a) curve has been
  marked in the graph with the corresponding channel number by a black
  dot. The inset shows the distribution for $N=2$ and its derivative.}
\label{fig:rhoic}
\end{figure}

The distribution functions for the full range of $i_c$ are generated
numerically via a Monte Carlo integration. The results are displayed
in Fig.~\ref{fig:rhoic}. The numerical error is of the order of the
line width. The distribution for $N=2$ again is shown in the inset
together with its derivative. The derivative indicates clearly
the feature at $I_c=e\Delta/\hbar$. For larger channel number $N$, the
distribution functions approach a Gaussian distribution. In the limit
$N\rightarrow\infty$, the Gaussian is characterized by $\langle I_c
\rangle = 0.35\,N e \Delta/\hbar$ and $\mbox{rms}\, I_c = 0.30\, e
\Delta/\hbar$, and the critical phase, at which the supercurrent
attains its maximum value, is $\phi_c = 2.04$. For $N=10$ we are
already close to this limit, with first moments $\langle I_c \rangle =
0.33 \pm 0.01\, N e \Delta / \hbar$ and $\mbox{rms}\, I_c = 0.29 \pm
0.05\, e \Delta / \hbar$.

\section{SUMMARY}
The distribution function for the dimensionless critical current
$i_c=I_c/( e\Delta/\hbar)$ through a chaotic quantum dot is, for a
large number of channels $N$, a Gaussian.\cite{beenakker92} Its
position is the average critical current of order $N$ and its width
are the root-mean-square fluctuations of $i_c$ of order unity.  With
decreasing channel number these two first moments of the distribution
function are still of order $N$ and 1, respectively, however the
prefactors change. Both, the average value of the critical current,
normalized to $N$, and the rms fluctuations of the critical current
decrease. This is not unexpected since the ``level repulsion'' between
the transmission eigenvalues of different channels declines.

In the extreme case of a purely one-dimensional normal conducting
contact, that is a single-channel contact, level repulsion is entirely
absent and for the COE the distribution function displays a
square-root divergence. This originates in the singular
behavior of the distribution function for the transmission eigenvalues
for the COE. Moreover, a general relation was established which
connects the distribution of transmission eigenvalues to the
distribution function for the critical supercurrent in the single
channel situation. For multiple channels we provided a small current
expansion of this distribution function which relates the number of
channels to the exponent of the algebraic small-current behavior. With
these relations it is in principle possible to identify the ensemble
by a measurement of the distribution of critical currents.

\section*{Acknowledgements}

We acknowledge many helpful discussions with J.C.~Cuevas,
A.~Mildenberger, A.D.~Mirlin, A.~Rosch and E.~Scheer. Special thanks
go to P.~W\"olfle who proposed the numerical determination of the
distribution of the critical current. We dedicate this paper with
great pleasure to Peter W\"olfle on the occasion of his 60th birthday.
This work was supported by the SFB 195 and by the BMBF 13N6918/1.

\section*{APPENDIX A}

Here, we elaborate on the determination of the spectral current $j(E,\phi)$, 
Eq.~(\ref{spectral-current}), for the single channel situation. An understanding 
of the energy dependence in this most simple case is useful for further discussions.
The typical energy scale for variations of the scattering matrix elements is the Thouless energy.

The scattering matrix $\mathcal{S}$ for the normal region of the contact is a $2 \times 2$ 
matrix. In the polar decomposition (\ref{polDekomp}) it is characterized by four phases 
$\theta_L, \theta_R, \chi_L, \chi_R$ and the transmission coefficient $T$ of this single
channel:
\begin{equation}\label{EinkanalS}
\mathcal{S} = 
\left( \begin{array}{cc} e^{i\theta_L} & 0 \\ 0 & e^{i\chi_L} \end{array}\right) 
\left( \begin{array}{cc} -\sqrt{1-T} & \sqrt{T} \\ 
                        \sqrt{T} & \sqrt{1-T} \end{array}\right) 
 \left( \begin{array}{cc} e^{i\theta_R} & 0 \\ 0 & e^{i\chi_R} \end{array}\right)
\end{equation}
In general, these five parameters of the scattering matrix are energy dependent.
We define the new phase parameters  $\theta = \theta_L+\theta_R$,
$\Delta\theta = \theta_L-\theta_R $, $\chi = \chi_L+\chi_R$ and $\Delta\chi =
\chi_L-\chi_R$
and find for the spectral current
\begin{eqnarray}\label{EinkanalG}
j(E,\phi)\!\!\!\!\!&=\!\!\!\!\!& -\frac{d}{d\phi}\Im\ln [ 
\cos(\varphi_1(E)\!+\varphi_2(E)\!-\frac{1}{2}(\theta(E)\!-\theta(-E)\!+\chi(E)\!-\chi(-E)) )  \nonumber\\ 
&&\!\!\!-\bar{R}(E)
\cos(\varphi_1(E)-\varphi_2(E)-\frac{1}{2}(\theta(E)-\theta(-E)-\chi(E)+
 \chi(-E)) )
\nonumber\\
&&\!\!\! - \bar{T}(E) \cos(\phi+
\frac{1}{2}(\Delta\theta(E)+\Delta\theta(-E)-\Delta\chi(E)-\Delta\chi(-E))) ] 
\end{eqnarray}
with the particle-hole symmetric values for transmission and reflection coefficients
\begin{eqnarray}
\bar{T}(E) &=& \sqrt{T(E)T(-E)} \nonumber\\
\bar{R}(E) &=& \sqrt{(1-T(E))(1-T(-E))} \,.
\end{eqnarray}

If the matrix $\mathcal{S}$ is only weakly energy dependent (in
particular no resonances should exist in an energy range of $2\Delta$
around the Fermi energy $E_F$) the above expression can be expanded:
according to the Andreev approximation the phases are taken to first
order in $E/E_F$ and the transmission coefficient in zeroth order.
This results in the low-energy expression for the spectral current
\begin{eqnarray}\label{AndreevEinkanalG}
j(E,\phi) &=& - \frac{d}{d\phi}\Im\ln [ 
\cos(\varphi_1(E)+\varphi_2(E)-(\theta'(0)+\chi'(0))E)  \nonumber\\ 
&& - R \cos(\varphi_1(E)-\varphi_2(E)-(\theta'(0)-\chi'(0))E)
\nonumber\\
&& - T \cos(\phi+\Delta\theta(0)-\Delta\chi(0)) ]
\end{eqnarray}
where the prime on the phases denotes a derivative with respect to energy, and
$R=1-T$. 

Only in the presence of time-reversal symmetry the phases
$\Delta\theta(0)$ and $\Delta\chi(0)$ vanish. However, in the absence
of time-reversal symmetry the whole $I(\phi)$ curve is shifted by
$\Delta\theta(0)-\Delta\chi(0)$. The mean supercurrent at a fixed
phase $\phi$ then requires averaging over the constant phase shift
which results in a vanishing supercurrent. On the other hand, a
measurement of only the critical current is not affected by this phase
shift and therefore expression (\ref{Distrib-Ic}) is valid
irrespective of the presence or absence of time-reversal symmetry.

\begin{figure}
%\vspace*{1cm}
\begin{center}
   \psfrag{0}{\small $0$} 
   \psfrag{0.1}{\small $0.1$}
   \psfrag{0.2}{\small $0.2$} 
   \psfrag{0.3}{\small $0.3$}
   \psfrag{0.4}{\small $0.4$}
   \psfrag{0.6}{\small $0.6$}
   \psfrag{0.8}{\small $0.8$}
   \psfrag{1}{\small $1$}
   \psfrag{COE}{\small COE}
   \psfrag{ballistic}{\small ballistic}
   \psfrag{d}{\small $\delta$}
   \psfrag{Ic-ball}{\small $I_c^{\mbox{\tiny{ballistic}}} \,\, [Ne\Delta_{\mbox{\tiny{max}}}/\hbar]$} 
   \psfrag{Ic-COE}{\small $\langle I_c \rangle_{\mbox{\tiny{COE}}} \,\, [Ne\Delta_{\mbox{\tiny{max}}}/\hbar]$}
     \leavevmode
\includegraphics[height=2.8in]{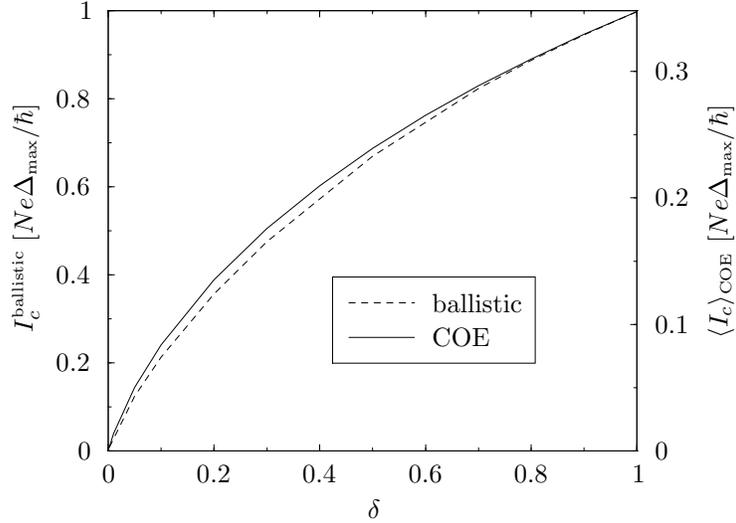}
\end{center}
\caption{Critical current for large $N$ of an asymmetric Josephson contact 
  averaged over the COE and of a ballistic contact for different
  ratios $\delta =
  \Delta_{\mbox{\tiny{min}}}/\Delta_{\mbox{\tiny{max}}}$. }
\label{fig:asymmetric}
\end{figure}

Assuming the transmission coefficients to be distributed according to
the COE, the averaged spectral current~(\ref{AndreevEinkanalG}) reads
in the presence of time-reversal symmetry
\begin{equation}\label{MittelEinkanalGreenCOE}
\langle j(E,\phi) \rangle_{\mbox{\tiny{COE}}} =
\Im\left\{\frac{\sin\phi}{\cos\phi - \cos(\varphi_1-\varphi_2-(\theta'-\chi')E)}\,
(1 - \frac{\arctan\sqrt{a}}{\sqrt{a}})\right\}
\end{equation} 
where
\begin{equation}\label{a-phi}
a = \frac{\cos(\varphi_1-\varphi_2-(\theta'-\chi')E) -\cos\phi}
{\cos(\varphi_1+\varphi_2-(\theta'+\chi')E) - \cos(\varphi_1-\varphi_2-(\theta'-\chi')E)}\,.
\end{equation}
The derivatives of the phases are taken at $E=0$. The averaged
spectral current (\ref{MittelEinkanalGreenCOE}) can be considered as a
basis for a more detailed evaluation respecting a weak energy
dependence of the scattering matrix.

If we neglect the energy dependence of the scattering matrix altogether, we arrive at
Eq.~(\ref{GreenF-zeroE}), which was the basic equation for the evaluation in the
section on single channel junctions.

\section*{APPENDIX B}

In the limit of a large number of channels $N$ the expression
(\ref{MultiChannelSpectCurrent}) for the spectral current of an
asymmetric Josephson junction can be averaged over the COE with the
distribution $\rho(T) = N/\pi\,[T(1-T)]^{-1}$. We get
\begin{eqnarray} \label{MittelVielkanalGreen}
\lefteqn{\langle j(E,\phi) \rangle =} \\
&&\Im\left\{\frac{N \sin\phi}{\cos\phi-\cos(\varphi_1\!-\!\varphi_2)}
\left(1- \left[1+\frac{\cos(\varphi_1\!-\!\varphi_2) - 
\cos\phi}{\cos(\varphi_1\!+\!\varphi_2)-\cos(\varphi_1\!-\!\varphi_2)}\right]^{-1/2} 
\right)\right\}\nonumber\,.
\end{eqnarray}
The averaged critical current $\langle I_c \rangle$ for large $N$ can
now be evaluated numerically and is shown in Fig.~\ref{fig:asymmetric}
versus the asymmetry parameter $\delta =
\Delta_{\mbox{\tiny{min}}}/\Delta_{\mbox{\tiny{max}}}$.  For $\delta =
1$ it starts at the value $\langle I_c \rangle = 0.35\, N e
\Delta/\hbar$ for the symmetric junction and decreases with decreasing
$\delta$. For comparison the values of $I_c$ for the ballistic
junction with $N$ open channels are also shown.


\begin{thebibliography}{100}
\bibitem{footnote1} A quantity $A=\sum_n a(T_n)$ is denoted a linear statistic
  on the transmission coefficients because products of $T_n$ are absent.
\end{thebibliography}
\end{document}